\documentclass{article}
\usepackage{amsmath}
\usepackage{amsthm}
\usepackage{amsfonts}
\usepackage{amssymb}
\usepackage{graphicx}
\usepackage{cmap}
\usepackage[T1]{fontenc}
\usepackage[utf8]{inputenc}
\usepackage[english]{babel}
\usepackage{authblk}
\usepackage{xcolor}

\newcommand{\ER}{{Erd\H{o}s-R\'enyi }}
 
\newcommand{\beq}{\begin{equation}}
\newcommand{\eeq}{\end{equation}}

\begin{document}
\title{Epidemic oscillations induced by social network control: the discontinuous case}

\author[1]{Daniele De Martino}
\author[2,3,4]{Fabio Caccioli}
\affil[1]{Biofisika Institute (CSIC,UPV-EHU) and Ikerbasque Basque Foundation for Science, Bilbao 48013, Spain }
\affil[2]{Department of Computer Science, University College London, Gower Street WC1E 6EA London (UK)}
\affil[3]{Systemic Risk Centre, London School of Economics and Political Sciences, WC2A 2AE, London (UK)}
\affil[4]{London Mathematical Laboratory, 8 Margravine Gardens, London WC 8RH (UK).}

\maketitle

\abstract{Epidemic spreading can be suppressed by the introduction of containment measures such as social distancing and lock downs. Yet, when such measures are relaxed, new epidemic waves and infection cycles may occur. Here we explore this issue in compartmentalized epidemic models on graphs in presence of a feedback between the infection state of the population and the structure of its social network for the case of discontinuous control. 
We show that in random graphs the effect of containment measures is simply captured by a renormalization of the effective infection rate that accounts for the change in the branching ratio of the network. In our simple setting, a piece-wise mean-field approximations can be used to derive analytical formulae for the number of epidemic waves and their length.  A variant of the model with imperfect information is used to model data of the recent covid-19 epidemics in the Basque Country and Lombardy, where we estimate the extent of social network disruption during lock downs and characterize the dynamical attractors.}

\section*{Introduction}

The onset of oscillations in a system as a consequence of feedback has been highlighted since the inception of control theory \cite{maxwell1868governors, andronov1966theory}.
Traditionally, feedback-induced oscillations have been studied in engineering artificial tools like thermostats and steering devices \cite{aastrom2010feedback}. More recently, research focused also on its application to natural systems \cite{pigolotti2007oscillation}, in particular homeostasis and its disruption in biological systems, a classical example being  glycemic control and diabetes in human metabolism \cite{frayn2009metabolic}.

Feedback-induced oscillations are currently emerging as governments are trying to control the evolution of the covid-19 pandemic crisis with containment measures such as social distancing, lock downs and quarantine. 

The modeling of containement measures in compartmentalized epidemic models \cite{murray2007mathematical} is  thus under the focus of intense research \cite{bianconi2020epidemics, dilauro2020epidemics}. It has been very recently rigorously demonstrated that compartimentalized epidemic models  display oscillations in presence of feedback between infection rate and infection states \cite{zhou2020active},and that in general a feedback between order and control parameters in large interacting systems subject to phase transitions triggers self oscillations \cite{de2018feedback, de2019oscillations}, where an Andronov-Hopf bifurcation takes over the usual phase transition.

As infection and recovery rates are changed, epidemic models on networks display out-of-equilibrium phase transitions between a phase where a disease is prevented from spreading and a phase where a macroscopic finite fraction of the population becomes infected \cite{hinrichsen2000non}.

In this article, we will study the SIS and SIR models in a full microscopic settings on random networks in presence of a feedback that changes the structure of the underlying social network, and we will show that such feedback triggers self-oscillations along the theory proposed in \cite{de2018feedback}, where suitably defined connectivity properties play the role of the control parameter. 
In order to mimic the occurrence of lock downs, we will focus on a a simple discontinuous feedback control, where a certain fraction of links is deleted if the fraction of infections exceeds a given threshold $I_2$. The same links are then be reinstated once the fraction of infections has been reduced below a second threshold value $I_1<I_2$.

Oscillations in epidemic spreading have been studied mainly from the point of view of seasonal effects that act as an external driving forces, while in the case studied here oscillations are autonomously driven by an internal feedback. The resulting models are described by time-independent equations and parameters, and such oscillations  can be considered  as emerging self-oscillations \cite{lauro2011self, buccheri2016experimental, jenkins2013self}. 

The article is organized as follows: In the first section we define the model and illustrate its behavior with results from numerical simulations on an instance of a real social network. In the second section we then study the model on mean-field uncorrelated networks, where we will show that the overall effect of lock downs on the dynamics is captured by a renormalization of the effective infection rate through a change of the network branching ratio. 

This finding is then exploited in the third section, where we analyze simple piece-wise well-mixed models with point transformation techniques, leading to analytical formulae for the number of waves and their length. In the last section, we consider the realistic case of imperfect information on the infection state, and we infer parameters from data on the current evolution of the covid-19 pandemic, for which we estimate the extent of social network disruption and characterize its limit cycle attractors.

\section*{Results}
\subsection*{Model definition}
We consider compartmentalized epidemic models on random networks, specifically the SIS and SIR models (see \cite{newman2018networks} for a review), where individual agents are represented as the nodes of a social network and can be in different states: Infected, Susceptible and Recovered. Infected individuals recover with a Poissonian rate $\gamma$, which is a parameter of the model, becoming either susceptible (in the SIS model) or recovered (in the SIR model), and they  infect neighboring susceptible nodes with a Poissonian rate $\beta$, which is the second parameter of the model. 
In order to model containment measures and their relaxation, we consider a feedback between the network structure and the infected state and its history in the following terms:
\begin{itemize}
\item Starting from a state with few infections (whose relative number we will indicate with $I$) that fastly spread, if the spreading overcomes a certain threshold $I>I_2$  a central authority decides to disrupt the network structure by randomly removing a macroscopic fraction $q$ of the links. This will eventually revert back the spreading.
\item Starting from a regressing  infected state in a disrupted network, when the infection state is reverted to an acceptably low value $I<I_1$, the network structure is restored back to its initial conditions. 
\end{itemize} 
\begin{figure}[h!!!!!]\label{fig1}
\begin{center}
\includegraphics*[width=0.9\textwidth,angle=0]{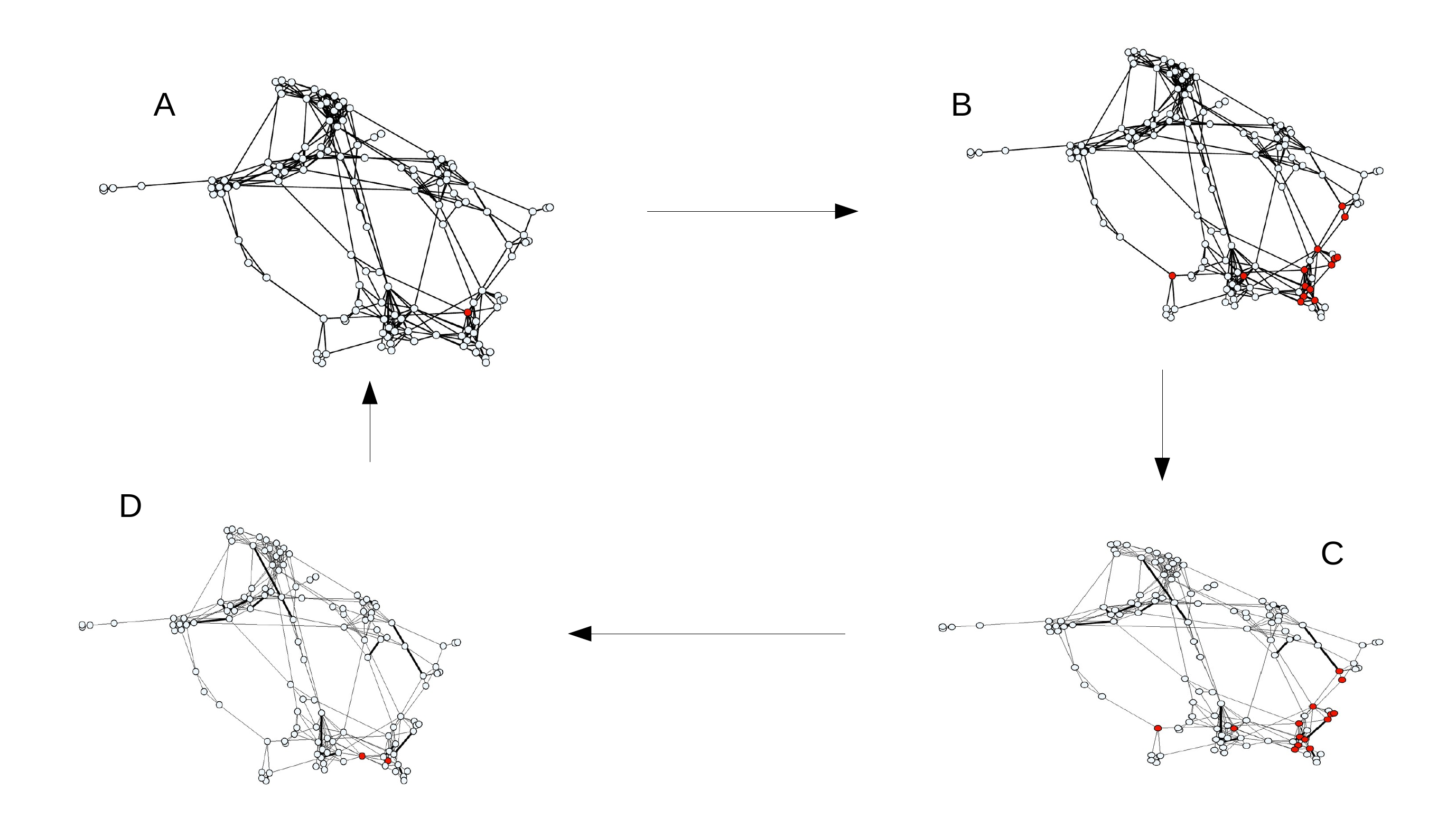}
\caption{\footnotesize\it $A \to B$: In a dense social network few infected highly contagious people give rise to an  epidemic spreading. $B \to C$: during the epidemic outbreak a centralized authority decides for containment measures by severing the network. $C \to D$ Under confinement, the epidemic regresses to few  cases. $D \to A$ Once the epidemic is supposedly under control containment measures are withdrawn and the social network is restored. Red dots represent infected individuals, while white dots represent susceptible ones. Light grey links in panels C and D represent links that have been removed because of the containment measure.}
\end{center}
\end{figure}
For the SIS model, this will eventually lead to an infection cycle as illustrated in Fig. 1, where we show results of simulations on a school friendship network reconstructed in \cite{mastrandrea2015contact} (number of nodes  $N=134$).

In Fig. 2, we show instead a simulation of the SIR model on the same network with and without feedback control, thus illustrating the effect of enforcing containment measures. The control successfully reduces the spreading of the infection, but for this to occur a series of lock downs have to be put in place.  

\begin{figure}[h!!!!!]\label{fig2}
\begin{center}
\includegraphics*[width=0.45\textwidth,angle=0]{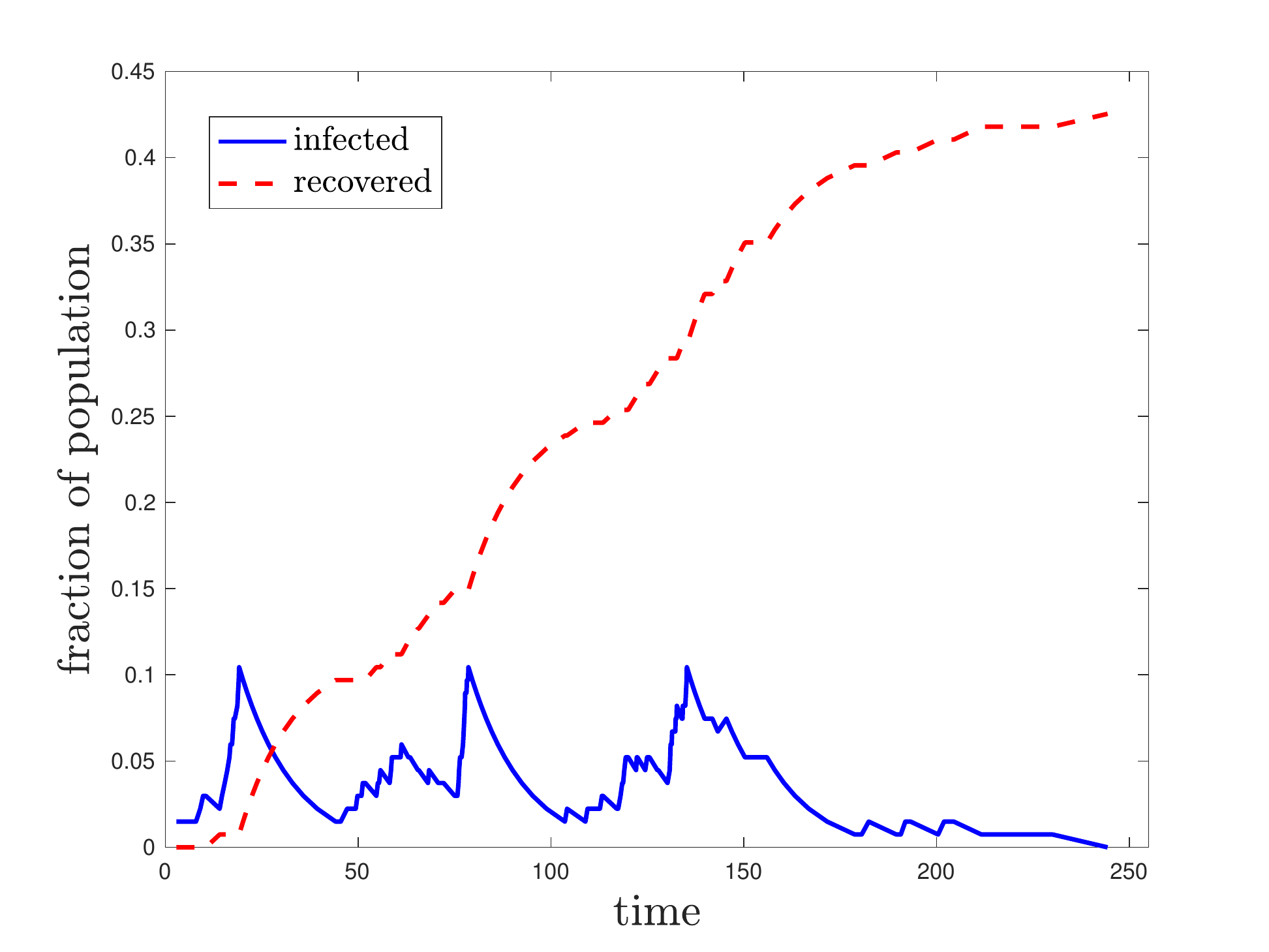}
\includegraphics*[width=0.45\textwidth,angle=0]{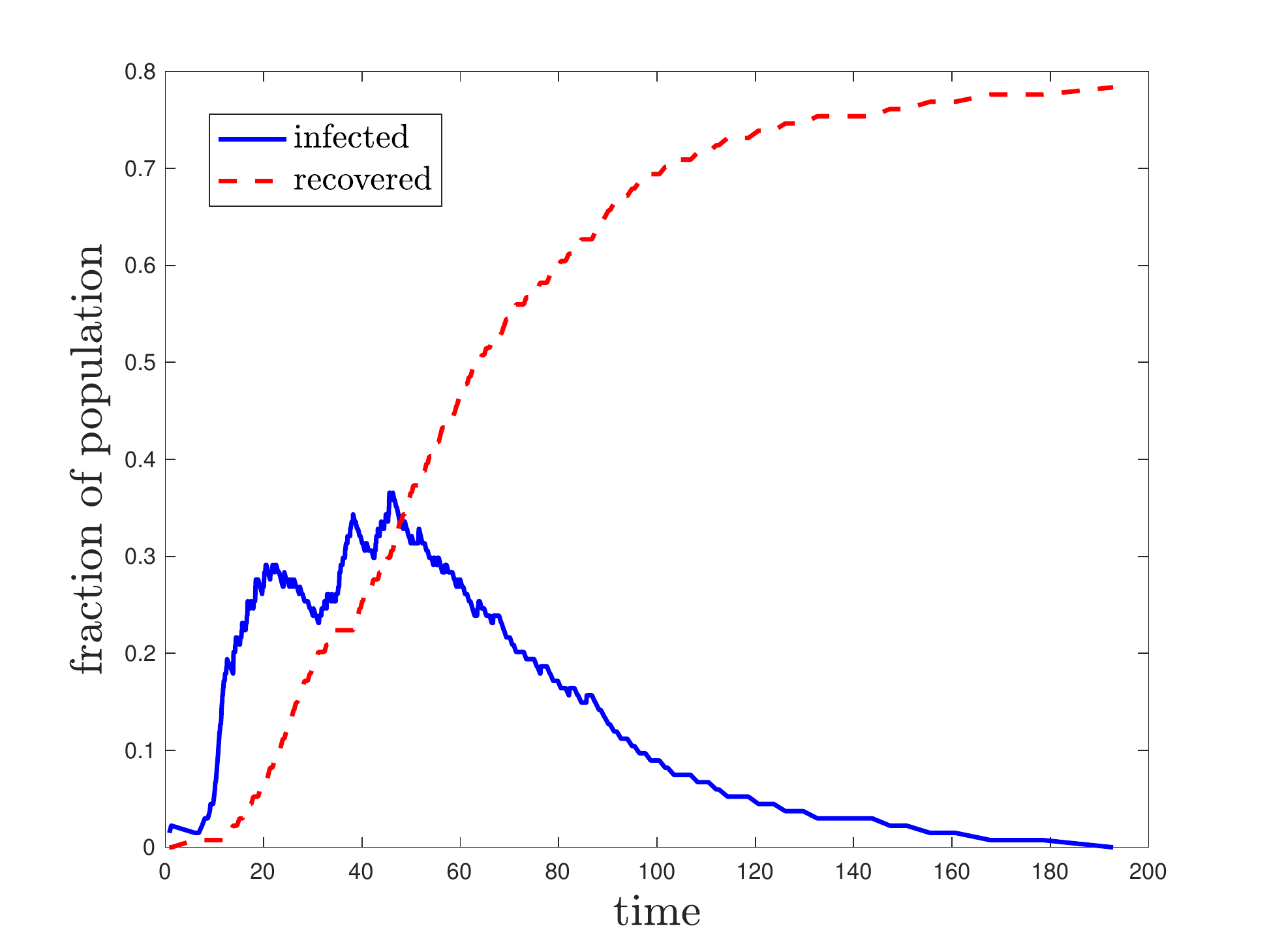}
\caption{\footnotesize\it Fraction of infected and subsequently recovered individuals as a function of time (that can be measured in days) from epidemic simulations of the SIR model in a school friendship network \cite{mastrandrea2015contact} (number of nodes  $N=134$)  with (left) and without (right) lock down measures in place, with parameters  $\beta=0.05$ (bare infection rate), $\gamma=0.07$ (recovery rate). 
The lock down for the feedback case is enforced if the number of infections is above $13$ and it is relaxed if they are below $2$, and it consists in a dilution of the network links by a factor $q=0.99$. There are three lock downs occurring at times  $t=25$, $t=80$ and $t=130$.}
\end{center}
\end{figure}

We will consider in the next section the SIS and SIR models for the case of large uncorrelated random networks, where it is possible to characterize analytically general features of the dynamics.  

\subsection*{Networks}

We consider here the case of large annealed uncorrelated random graphs with degree distribution $P(k)$. At odds with static networks, in annealed networks we assume that links are randomly rewired over a faster time scale of the spreading process, while the assumption of uncorrelated networks implies there are no correlations between the degrees of neighboring nodes.
The locally treee-like structure of these networks makes it possible to make analytical progress in the study of dynamical processes taking place on them, since it allows to recur to well-controlled approximations for the factorization of probability states. In particular, we will consider here the heterogeneous mean-field approximation, where nodes are grouped in classes according to their degree. 

Let us start from the SIS model. In absence of feedback, the rate equation for the fraction $I_k$ of infected individuals of degree $k$ can be written as follows \cite{pastor2015epidemic}
\beq
\dot I_k(t) = \beta
\left(1-I_k(t)\right) k \Theta(t) -\gamma I_k(t),
\eeq
where $\Theta(t)=\sum_k\frac{kP(k)}{\langle k\rangle} I_k(t)$ is the probability that a randomly selected neighbor of a node of degree $k$ is infected, and we denote by $\langle k\rangle$ the average degree of the network. 

Here we consider the case in which, when the fraction of infected individuals $I(t)=\sum_k P(k) I_k(t)$ exceeds a given threshold $I_2$, a lock down measure is implemented that removes a fraction $q$ of links, which are then reinstated once the condition $I(t)<I_1$ is satisfied. The equations of the model in presence of this feedback mechanism can therefore be written in terms of a state-dependent infection rate as follows

\beq
\dot I_k(t) = \tilde\beta(I,\dot I) \left(1-I_k(t)\right) k \Theta(t) -\gamma I_k(t), \label{HMF_SIS}
\eeq

where

\beq
\tilde\beta(I,\dot I) = \begin{cases} 
\beta, {\rm if}~I(t)<I_1~{\rm or}~\begin{cases} I_1\le I(t)\le I_2\\ \dot I(t) >0\end{cases} \\
(1-q)\beta,~~{\rm if}~I(t)>I_2~{\rm or}\begin{cases}I_1\le I(t)\le I_2\\ \dot I(t)<0\end{cases}.
\end{cases}\eeq

We note that we can express $\tilde\beta$ as a function of the fraction of infected population and its derivative because we are considering deterministic rate equations. A more general representation for the discrete stochastic case would require the introduction of a binary state variable to denote the occurrence or absence of a lock down.

In Figure 3, we compare the result of numerical simulations with the numerical solution of equation \eqref{HMF_SIS} for \ER and scale-free random networks. In both cases we clearly see the emergence of oscillations due to the feedback. 

\begin{figure}[h!!!!!]
\begin{center}
\includegraphics*[width=0.45\textwidth,angle=0]{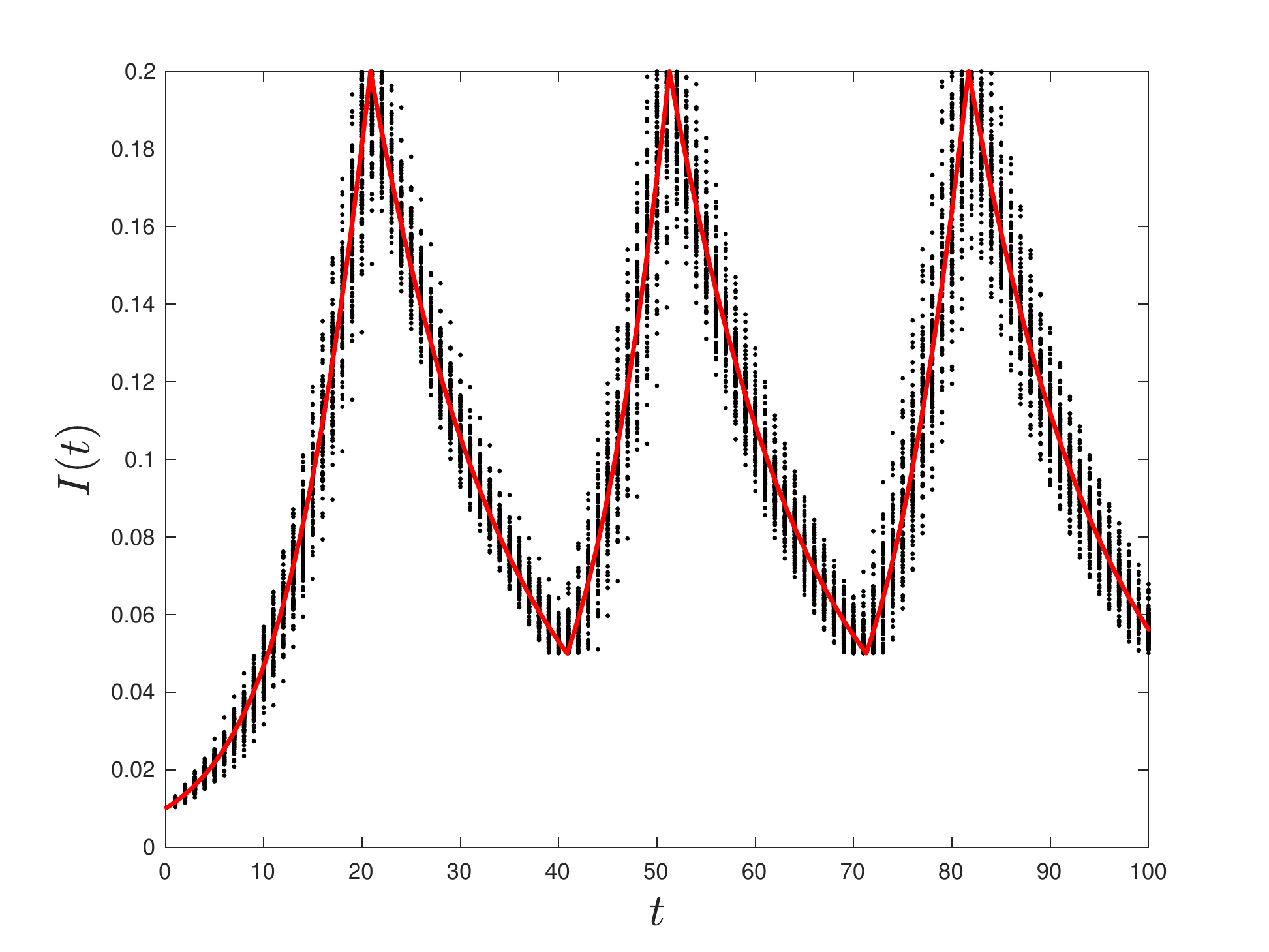}
\includegraphics*[width=0.45\textwidth,angle=0]{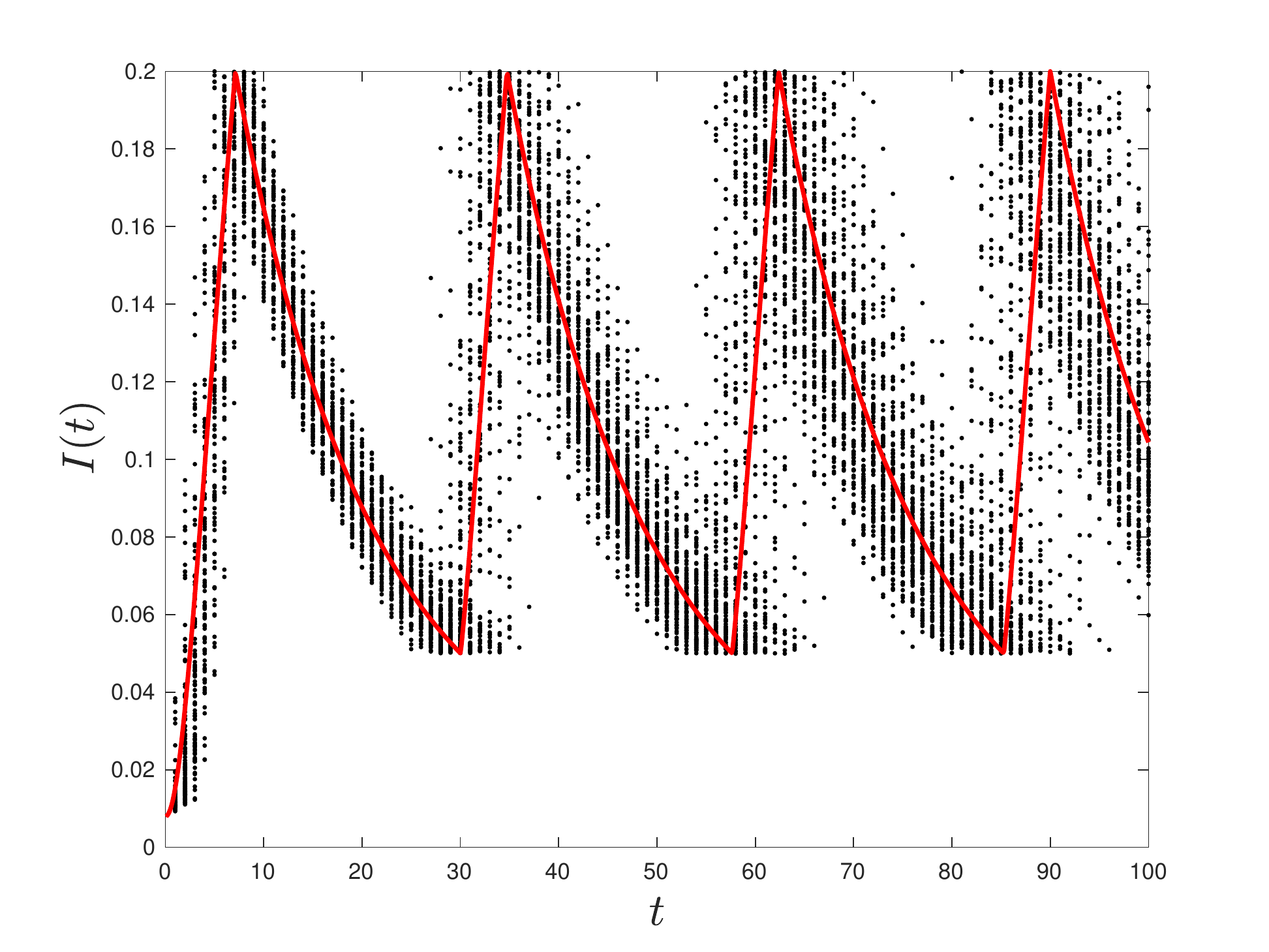}
\caption{\footnotesize\it Fraction of infected individuals as a function of time in an homogeneous and heterogeneous network (both with $N=10^5$ nodes ). Dots refer to 100 simulations of the SIS model (parameters $\beta=0.03$ $\gamma=0.08$ $I_1=0.05$ $I_2=0.2$ $q=0.95$), solid lines refer to the numerical integration of the mean-field equations . Left: \ER random network of average degree $\langle k \rangle=8$.  Right: scale-free network of minimum degree $ k_{min} =2$ and exponent $\alpha=2.5$. 
}
\end{center}
\label{fig_SIS}
\end{figure}

In Figure 4, we visualize the feedback-induced oscillations by means of a phase portrait, where we plot the fraction of infected individuals vs. the fraction of new positives. The figure clearly shows the emergence of a limit cycle as an attractor of the dynamic.
We also note from both Figures 3 and 4 that the dynamics on scale-free networks display bigger sample-to-sample fluctuations than that on \ER networks.

\begin{figure}[h!!!!!]
\begin{center}
\includegraphics*[width=0.45\textwidth,angle=0]{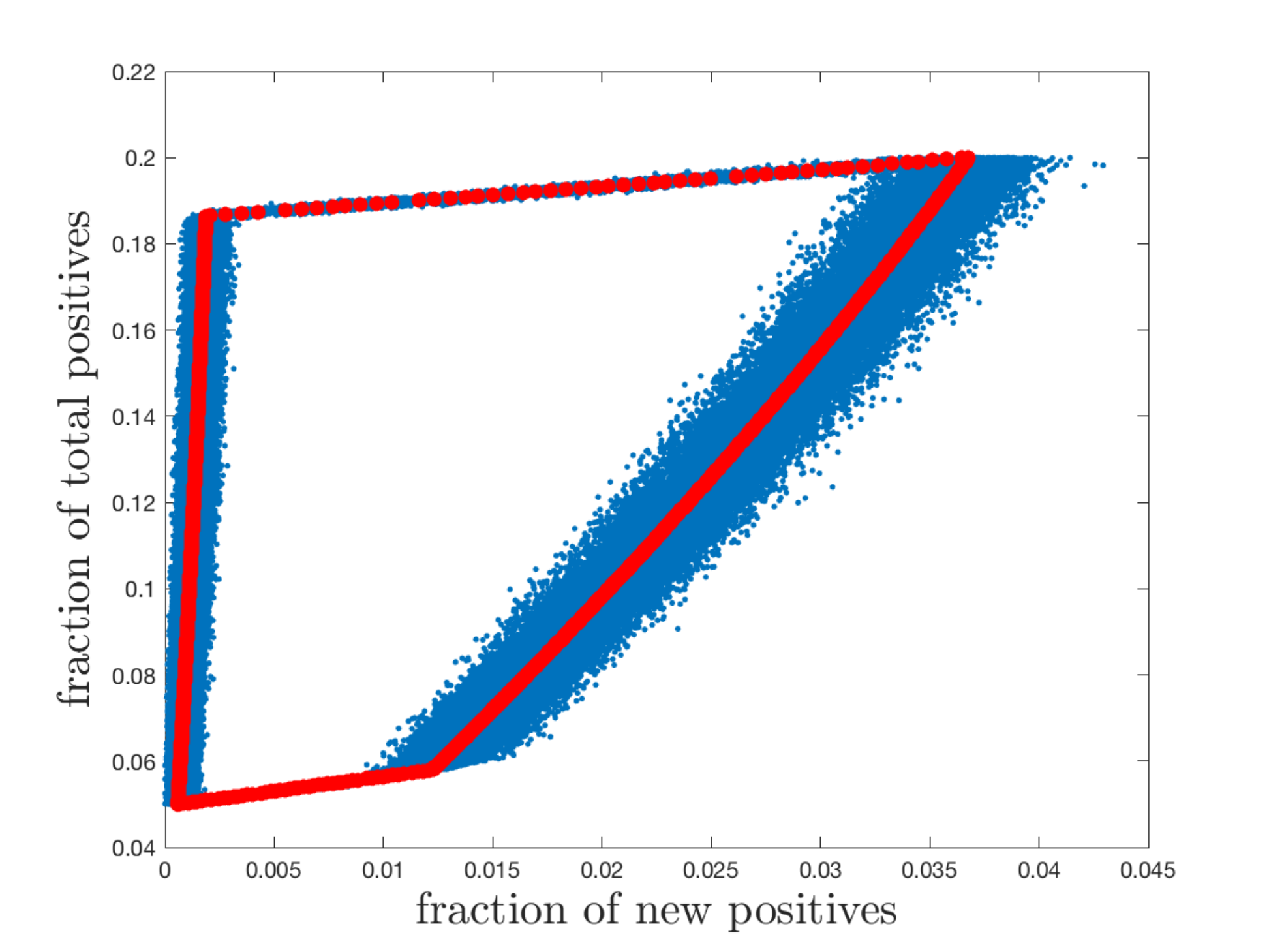}
\includegraphics*[width=0.45\textwidth,angle=0]{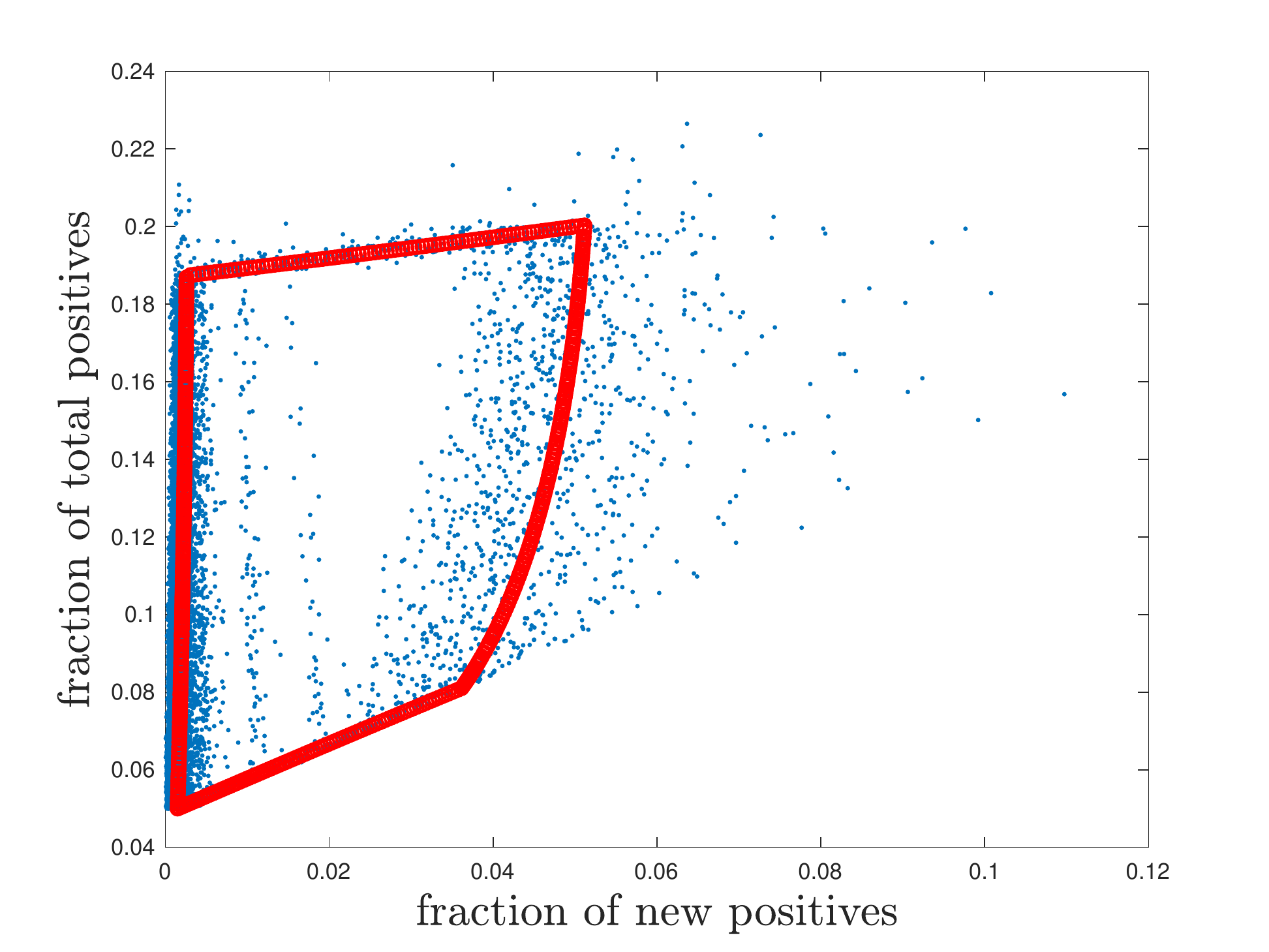}
\caption{\footnotesize\it Left panel: Phase portrait of an SIS model in an \ER network with $N=10^5$ and average degree $\langle k \rangle=8$. Right panel: Phase portrait of an SIS model in a scale-free network with $N=10^5$, $\gamma=2.5$  $k_{\rm min}=2$. In both panels, blue dots refer to 100 simulations of the SIS model (parameters $\beta=0.03$ $\gamma=0.08$ $I_1=0.05$ $I_2=0.2$ $q=0.95$). Red dots refer to the numerical integration of the mean-field equations.}
\end{center}
\label{fig_SIS}
\end{figure}

\begin{figure}[h!!!!!]
\begin{center}
\includegraphics*[width=0.45\textwidth,angle=0]{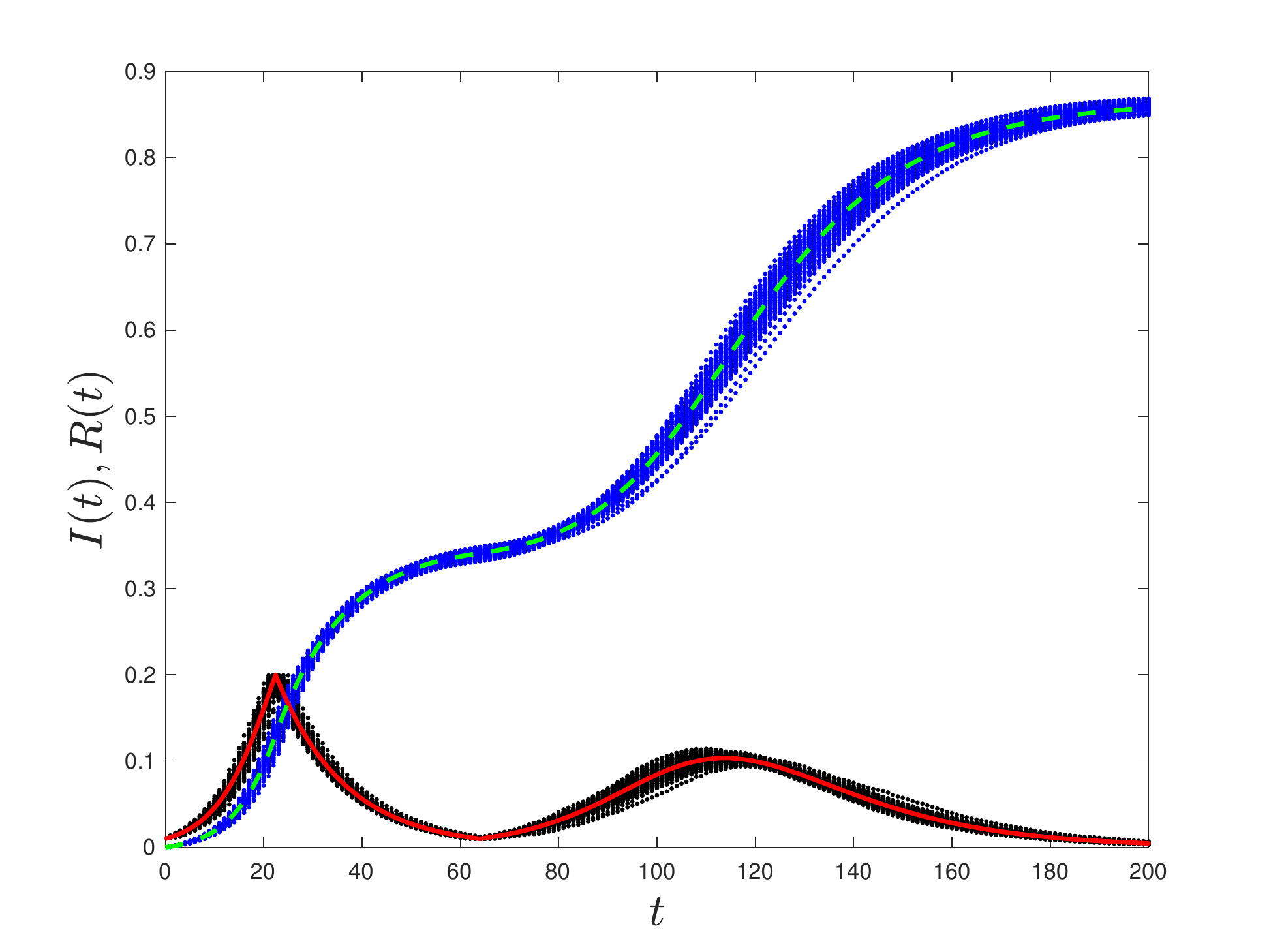}
\includegraphics*[width=0.45\textwidth,angle=0]{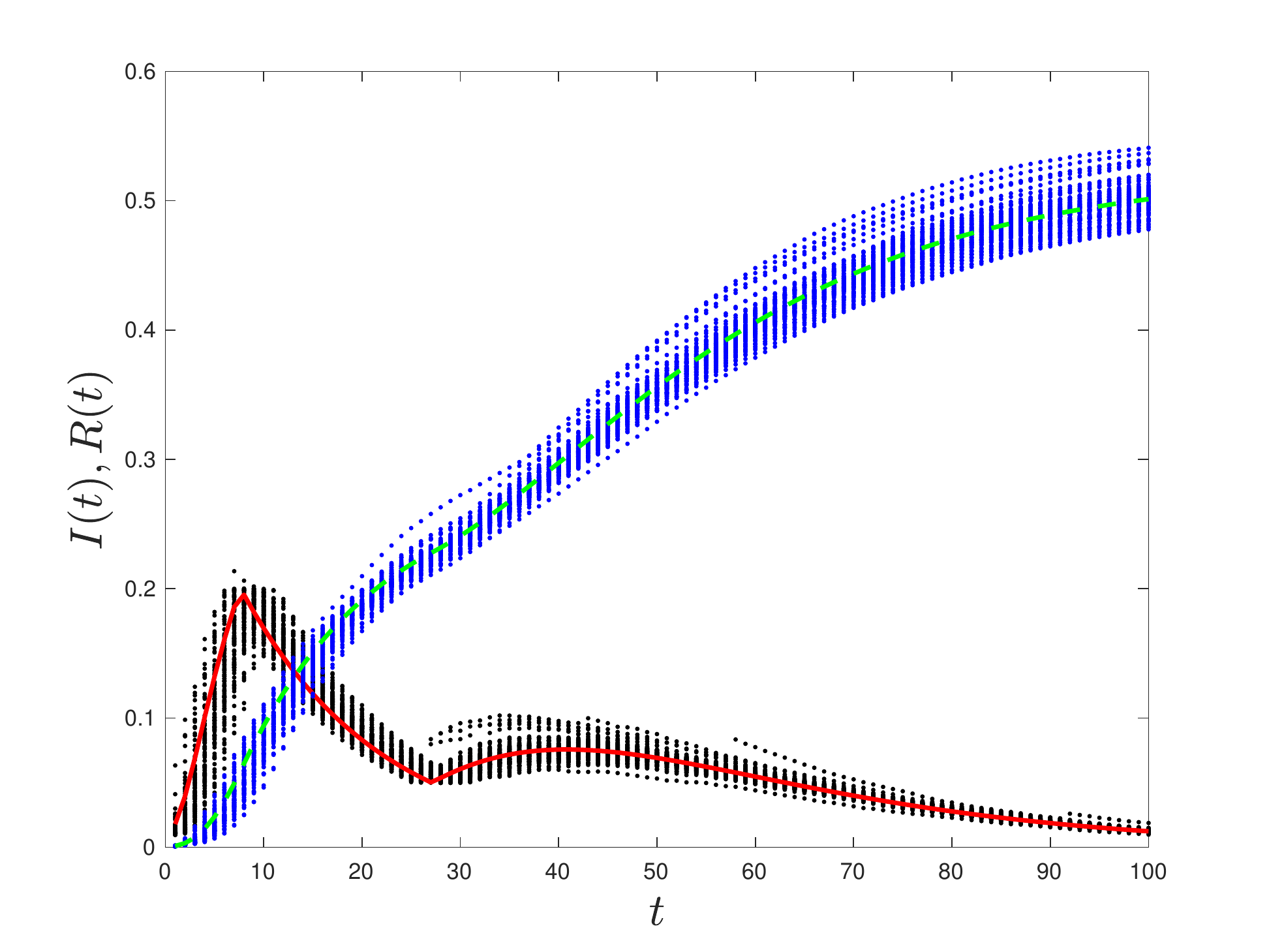}
\caption{\footnotesize\it Fraction of infected and recovered individuals as a function of time for the SIR model with feedback. Left panel: \ER network with $N=10^5$ nodes. Data refer to 100 simulations with parameters $\beta=0.03$ $\gamma=0.08$ $I_1=0.2$ $I_2=0.01$ $q=0.95$. Right panel: scale-free network with $N=10^5$ nodes, $\gamma=2.5$ and $k_min=2$.  Data refer to 100 simulations with parameters $\beta=0.03$ $\gamma=0.08$ $I_1=0.2$ $I_2=0.05$ $q=0.95$. Black points: fraction of infected individuals from numerical simulations. Blue points: fraction of recovered individuals from numerical simulations. Solid red line: fraction of infected individuals from numerical integration of the mean-field equations. Dashed green line: fraction of recovered individuals from numerical integration of the mean-field equations. 
}
\end{center}
\label{fig_SIR}
\end{figure}

The same feedback mechanism can be considered for the SIR model as well. If we now define $R_k$ as the fraction of recovered nodes with degree $k$, the rate equations that describe the dynamic of the SIR model are

\begin{eqnarray}
\dot I_k(t) &=& \tilde \beta \left(1-I_k(t)-R_k(t)\right) k \Theta(t) -\gamma I_k(t)\\
\dot R_k(t) &=& \gamma I_k(t),
\end{eqnarray}

\noindent where, as before, $\Theta(t)=\sum_k\frac{kP(k)}{\langle k\rangle} I_k(t)$\footnote{We are considering here the case of annealed networks, where links are randomly rewired at each time step. In the case of a static network, the factor $k$ in the definition of $\Theta$ would be replaced by a factor $k-1$.} and $\tilde\beta (I,\dot I)$ is given as before by equation (3).

In figure 5 we show the evolution over time of the fraction of infected and recovered individuals for the case of \ER and scale-free networks. We see that the introduction of the feedback can lead to oscillations corresponding to multiple infection waves. Clearly, in contrast to the case of the SIS model, these oscillations will eventually come to an end once a large enough fraction of the population has been infected. The number of infection waves  depends on the parameters of the model. In the next section we provide an analytical estimation for a well-mixed population in the limit when $q$ is close to $1$.

\subsection*{Mixed approximation}
The analysis reported in the previous section, where the effect of network embedding and the feedback enacting on it is captured by a simple renormalization of the bare infection rate, suggests to study the model on well-mixed populations to get analytical insights. In this section $I$, $S$ and $R$ will indicate the fraction of infected, susceptible and recovered individuals in the total population, respectively. 
The feedback law mimicking containment, with parameters $\beta_1>\beta_0$ and $I_2>I_1$, is given by
\beq\label{betaEff}
\tilde\beta(I,\dot I) = \begin{cases} 
\beta_1, {\rm if}~I(t)<I_1~{\rm or}~\begin{cases} I_1\le I(t)\le I_2\\ \dot I(t) >0\end{cases} \\
\beta_0,~~{\rm if}~I(t)>I_2~{\rm or}\begin{cases}I_1\le I(t)\le I_2\\ \dot I(t)<0\end{cases}.
\end{cases}\eeq

For this form of the function $\tilde\beta(I,\dot I)$, mixed epidemic models can be readily analytically solved piece-wise in each sector, where the infection rate is constant, and the solutions can be joined at the boundaries (see Andronov \cite{andronov1966theory}).   

For instance, in the mixed approximation the SIS model with no feedback and infection rate $\beta$ has the simple solution
($S+I=1$)
\begin{equation}
I_\beta (t) = \frac{1-\gamma/\beta}{\left(\frac{1-\gamma/\beta}{I(0)}-1\right)e^{-(\beta-\gamma)t}+1}  
\end{equation}
In presence of the feedback, the piece-wise constructed solution shows that for $\beta_1>\gamma>\beta_0$, $I_2<1-\gamma/\beta_1$ the dynamics settles into a limit cycle, and the periods of the quiescent  epidemic spreading ($t_1$) and of the recovery under lock downs ($t_2$) are given by the following analytical formulae
\begin{eqnarray}
t_1 = \frac{1}{\beta_1-\gamma}   \log \left( \frac{\frac{1-\gamma/\beta_1}{I_1}-1}{ \frac{1-\gamma/\beta_1}{I_2}-1} \right) \\
t_2 = \frac{1}{\beta_0-\gamma}   \log \left( \frac{\frac{1-\gamma/\beta_0}{I_2}-1}{\frac{1-\gamma/\beta_0}{I_1}-1} \right)
\end{eqnarray}
For a swift and resolute population lock down, we can  approximate $I_2 << 1-\gamma/\beta_1$, $\beta_0<<\gamma$ and obtain for the total duration of an epidemic wave 
\begin{equation}
T = t_1 + t_2 \sim \frac{\log(I_2/I_1)}{\gamma}\frac{R_0}{R_0-1}.
\end{equation}
where $R_0 = \beta_1/\gamma$.
For instance, from the values $R_0\sim 3$, , $I_2/I_1 \sim 100$ and $1/\gamma \sim 2$ weeks we can calculate  $T\sim 3$ months.

For the SIR model, the solution in each interval -- starting from initial conditions $S_i,R_i,I_i$ at time $t_i$ -- reads
\begin{eqnarray}
I+R+S &=& 1 \\
S &=& S_i e^{\beta/\gamma (R_i-R)} \\
t-t_i &=& \int_{R_i}^R  \frac{dr}{1-r-S_i e^{\beta/\gamma (R_i-r)}}.
\end{eqnarray}
The total number of lock downs can be worked out analytically by joining solutions piece-wisely, and a first order expansion in $(\beta_0/\beta_1, \beta_0/\gamma)$ (see the appendix) gives the formula
\begin{equation}
n^* (\beta_0) \sim n^* (\beta_0=0)/x, 
\end{equation}
where
\begin{equation}
n^* (0) = \frac{1-(1+\log R_0)/R_0}{I_2-I_1} 
\end{equation}
and
\begin{equation}
1-\beta_0/\beta_1\leq x \leq 1+\beta_0/\beta_1 + \beta_0/\gamma. 
\end{equation}
For instance for the values $R_0\sim 3$, $I_2-I_1 \sim 0.1 \pm 0.05$ we get 
$n^* \sim 3 \pm 2$. 
In the next section we will illustrate data modeling applications of our framework.

\subsection*{Imperfect information and data modeling}
In this section, we illustrate our framework in the context of modeling epidemic data of the covid-19 infection in 2020 in the Italian region of Lombardy and the Spanish region of the Basque Country. Data includes daily reports of new infections and active cases \footnote{From {\it https://github.com/pcm-dpc/COVID-19} (Lombardy) and \\  {\it https://opendata.euskadi.eus/catalogo-datos} (Basque country)}, plus a single prevalence estimate.
For the purpose of data analysis, we consider a variant of the model with imperfect information by splitting the total number of infections into detected and undetected cases, whose numbers we denote by $I_d$ and $I_u$ respectively.
We assume the existence of a detection process by which undetected infected individuals are spotted with rate $r$ and then put in isolation (which is equivalent to removing them). 
The mean-field rate equations are as follows
\begin{eqnarray}
\dot{I}_u &=& \tilde \beta(I_{d},\dot{I_d}) I_u S -(\gamma+r) I_{u} \\
\dot{I}_d &=& r I_{u} -\gamma I_{d}
\end{eqnarray}
where $\tilde \beta(I_d, \dot{I_d})$ is the piece-wise constant function defined in equation \eqref{betaEff} (parametrized by $\beta_1,\beta_0,I_1,I_2$), and  analogous equations for $S$ and/or $R$, depending on whether we consider the SIS and/or the SIR model. 

The inference of the model parameters has been performed by approximately solving Bayes equations under the hypothesis of Gaussian noise by means of Monte Carlo methods and dynamical system numerical simulations (see the appendix). 
\begin{figure}[h!!!!!]
\begin{center}
\includegraphics*[width=0.45\textwidth,angle=0]{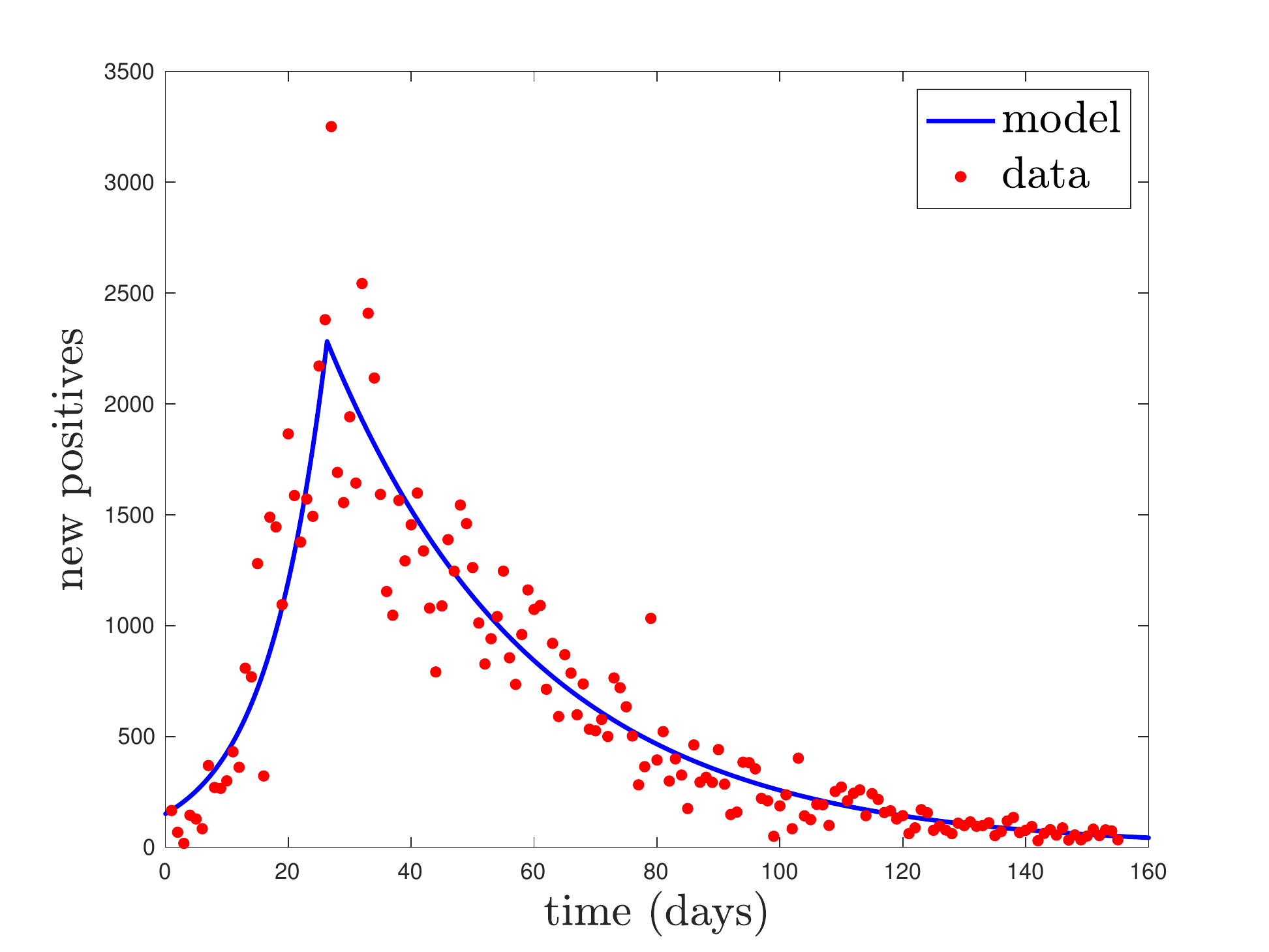}
\includegraphics*[width=0.45\textwidth,angle=0]{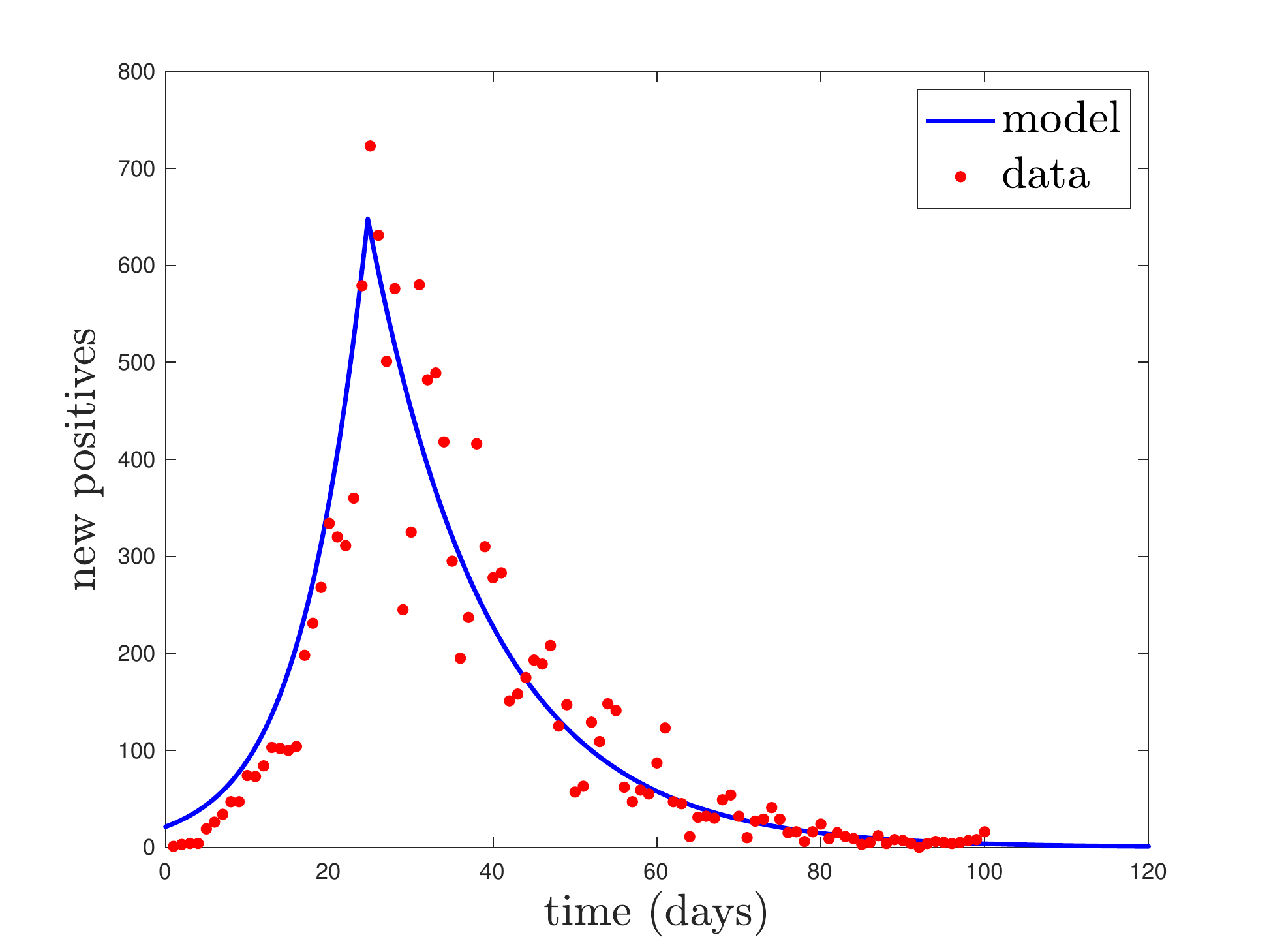}
\caption{\footnotesize\it New daily cases  vs time for the first epidemic wave of COVID-19: data from Lombardy (left) and Basque country (right) against the inferred model (maximum likelihood estimate).}
\end{center}
\end{figure}
\begin{figure}[h!!!!!]
\begin{center}
\includegraphics*[width=0.45\textwidth,angle=0]{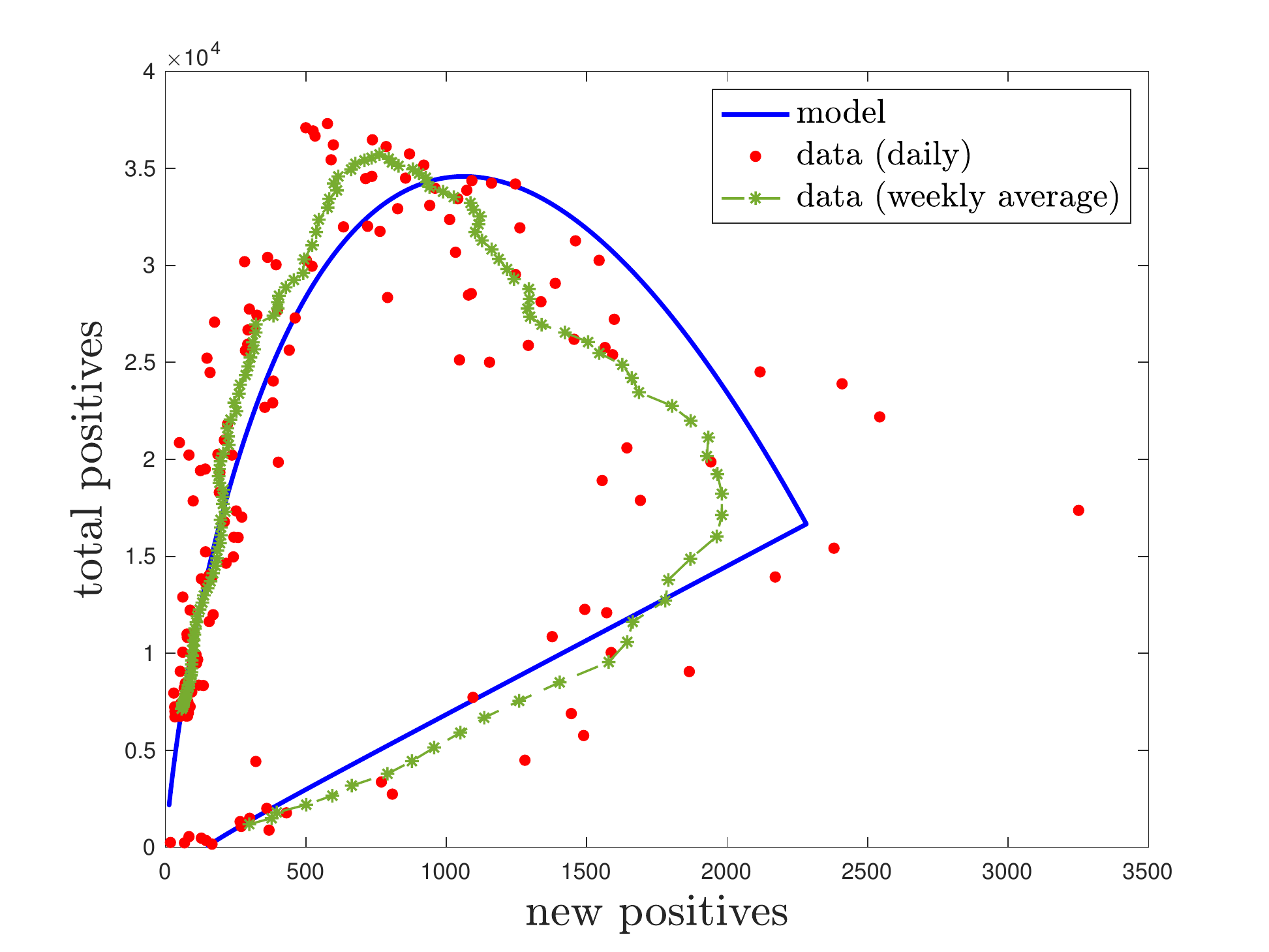}
\includegraphics*[width=0.45\textwidth,angle=0]{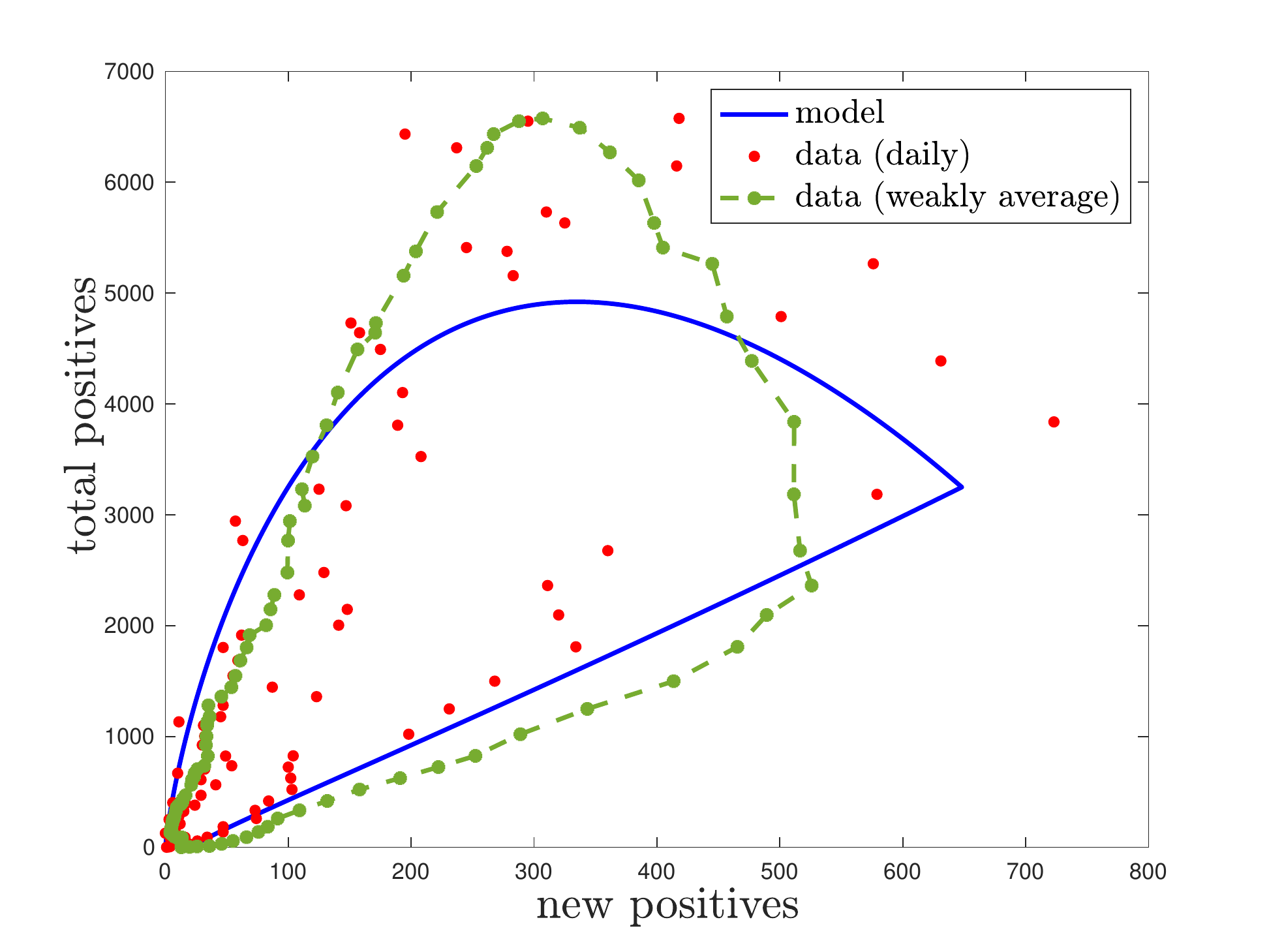}
\caption{\footnotesize\it Scatter plots of new daily and active cases for Lombardy (left) and Basque country (right) for the first epidemic wave of covid-19. In both panels, the solid blue line refers to the model inferred maximizing the likelihood, red dots refer to daily data and green stars refer to weekly data.}
\end{center}
\end{figure}

In Fig. 6 we show the time series of new daily cases for the two  analyzed cases against the model with maximum likelihood parameters, the latter showing a clear cusp peak and piece-wise exponential trends corresponding to the lock down event. The model suggests that a closed trajectory must be observed in the plane of new and active daily cases. This is shown in Fig. 7, where we report data smoothened by a 7-days moving average as well. This closed trajectory in the ideal case of a  pure SIS model with feedback would be a limit cycle attractor of the dynamics.

\section*{Conclusions}
Connectivity plays a crucial role in the definition of the parameters that control the collective behavior of a system.
This finding has striking consequences, like the absence of an epidemic threshold in epidemic spreading models defined on scale-free networks \cite{pastor2001epidemic}, and it suggests that it is possible to control the spreading by acting on the network of social interactions.

In this article we have shown that feedback control at the level of the social network in epidemic models triggers self-oscillations along the theory proposed in \cite{de2018feedback}.

We have investigated self-oscillations induced by a simple discontinuous feedback control mimicking lock down events in classical compartmentalized epidemic models (SIS and SIR) on networks. 

On random graphs, for \ER as well as scale-free networks with naive populations, we have shown that the effect of lock downs simply amounts at renormalizing the effective infection rate to account for the reduction in the network branching ratio.

This led to simple piece-wise mean field approximations that we solved analytically by means of transformation point methods, recovering formulae for the number of waves and their extent in terms of the model parameters. 
These formulae can be in principle tested against data, once a certain amount of evidence accumulates on the number of lock downs and their lengths, prevalence estimates, and basic infection numbers region by region. 

A problem related to data collection during the covid-19 epidemic outbreak was the fact that many positives were undetected. In order to bring the model to data, we have therefore extended it by assuming the existence of a fraction of undetected positives, who can then be detected at a given rate (for instance through testing).
We have applied our extended framework to analyze data from the first epidemic wave of covid-19 in Lombardy and Basque country,  where parameters have been inferred leading to a characterization of the dynamical attractors in the phase space.
Apart from applications to predictive modeling -- which would require more extensive data analysis \cite{vilar2020} and methods of system identification- - we do point out here briefly some potentially interesting theoretical problems stemming from this work.

First, the issue of optimal scheduling \cite{kirk2004optimal} in the control of the social network, leading to continuous feedback and potentially smoother oscillations. Current qualitative evidence from the second epidemic wave of covid-19 seems indeed to show in some regions smoother trends and oscillations around the phase transition point ($R_t \sim 1$), due to attempts of finer control like partial restrictions and selected closures taken in due course. Recently proposed analytical frameworks \cite{bianconi2020epidemics}  can be very useful in this respect.

Second, within the framework proposed here, where epidemics can be regarded as self-oscillators, it comes naturally the question of coupling and synchronization \cite{guerra2009coupled, di2018ginzburg} of epidemic waves running on different  networks that are weakly connected, e.g. by migration processes.
  
Finally, another interesting issue concerns the impact of periodically external drive on oscillators:  analogously to well-known forced double well oscillator \cite{guckenheimer2013nonlinear}, the combined effect of seasonal changes and feedback could potentially lead to chaotic oscillations in strange attractors, an aspect that adds to the problem of predictability of such systems, and that we leave for future investigations.

\appendix

\section{Number of epidemic waves: perturbative expansion}
In this section we derive formula (14-16). 
In the SIR model at fixed infection rate $\beta$ (as in a simple  model without feedback, or in a given interval for the piece-wise feedback model) the peak value of the infected fraction is (when $\dot{I_p}=0$ and $S_p= \gamma/\beta$)
\begin{equation}
I_p = I_i +S_i -\frac{\gamma}{\beta} \log S_i -\gamma/\beta (1-\log(\gamma/\beta)).
\end{equation}
This value is not achieved if it is greater than the one triggering the lock down, i.e. when $I_p > I2$. Thus we will assume as  halting condition that $I_p\leq I_2$, since in this case no lock down takes place and the system proceeds towards herd immunity. We will now work out a series for fraction of susceptible individuals at the various stages of the epidemic waves, exploiting the piece-wise analytical solutions in each interval.

Suppose we are at the beginning of a wave $I=I_1$ with given susceptible fraction $S=S_{n,-}$, the system (with $\beta=\beta_1$) will evolve towards $I_2$ and a given $S_{n,+}$  that satisfies
\begin{equation}
S_{n,+}-\frac{\gamma}{\beta_1} \log S_{n,+}=  S_{n,-}-\frac{\gamma}{\beta_1} \log S_{n,-} -\Delta I,
\end{equation}
where $\Delta I =I_2-I_1$.
Then we have the lock down $\beta=\beta_0$, and the system will evolve towards $I_1$ with a given $S_{n+1,-}$ that satisfies
\begin{equation}
S_{n+1,-}-\frac{\gamma}{\beta_0} \log S_{n+1,-}=  S_{n,+}-\frac{\gamma}{\beta_0} \log S_{n,+} +\Delta I
\end{equation}    
These equations define a series eventually halting when $I_p \leq I_2$.

\subsection*{Vanishing $\beta_0$}
Suppose  $\beta_0 = 0$.
In this case $S_{n+1,-}= S_{n,+} \equiv S_n$ and we have 
\begin{eqnarray}
S_{n+1}-\frac{\gamma}{\beta_1} \log S_{n+1} &=&  S_{n}-\frac{\gamma}{\beta_1} \log S_{n} -\Delta I \\
&=&  S_{n-1}-\frac{\gamma}{\beta_1} \log S_{n-1} -2 \Delta I  \\
&=&  \dots \\
&=& S_0-\frac{\gamma}{\beta_1} \log S_0 -n \Delta I
\end{eqnarray}
If we start from $S_0 \approx 1$, from the halting condition we find that the number  $n^*$ of lockdowns is 
\begin{equation}\label{nStarAppendix}
n^* = \left\lfloor\frac{1-(1+\log R_0)/R_0}{\Delta I}\right \rfloor,
\end{equation}
where we denote by $\left \lfloor x \right\rfloor$ the integer part of $x$.

\subsection*{First order expansion in $\beta_0$}

A first order expansion in $\beta_0$ leads  to
\begin{equation}
S_{n+1,-} \sim S_{n,+} (1-\beta_0/\gamma \Delta I).
\end{equation}
Then defining 
\begin{equation}
F_n = S_{n,-} -\gamma/\beta_1 \log S_{n,-}
\end{equation}
we have the recursion relation
\begin{equation}
F_{n+1} = F_n - x_n \Delta I  
\end{equation}
where
\begin{equation}
x_n =  1-\beta_0/\beta_1+\beta_0/\gamma S_{n+1,-}
\end{equation}
that can be bounded by $\mathcal{O}(\beta_0)$ terms (given that $ 0 \leq S_{n,-}\leq 1$)
\begin{equation}
 1-\beta_0/\beta_1  \leq x_n \leq  1-\beta_0/\beta_1 +\beta_0/\gamma
\end{equation} 
and the halting criterion leads to 
\begin{eqnarray}
n^* (\beta_0) \sim n^* (\beta_0=0)/x \\
1-\beta_0/\beta_1\leq x \leq 1-\beta_0/\beta_1 + \beta_0/\gamma 
\end{eqnarray}

In Figure 8 we show the agreement between numerical simulations on an \ER network and the analytical prediction given by equation \eqref{nStarAppendix} for $\beta_0=0$.

\begin{figure}[h!!!!!]
\begin{center}
\includegraphics*[width=0.7\textwidth,angle=0]{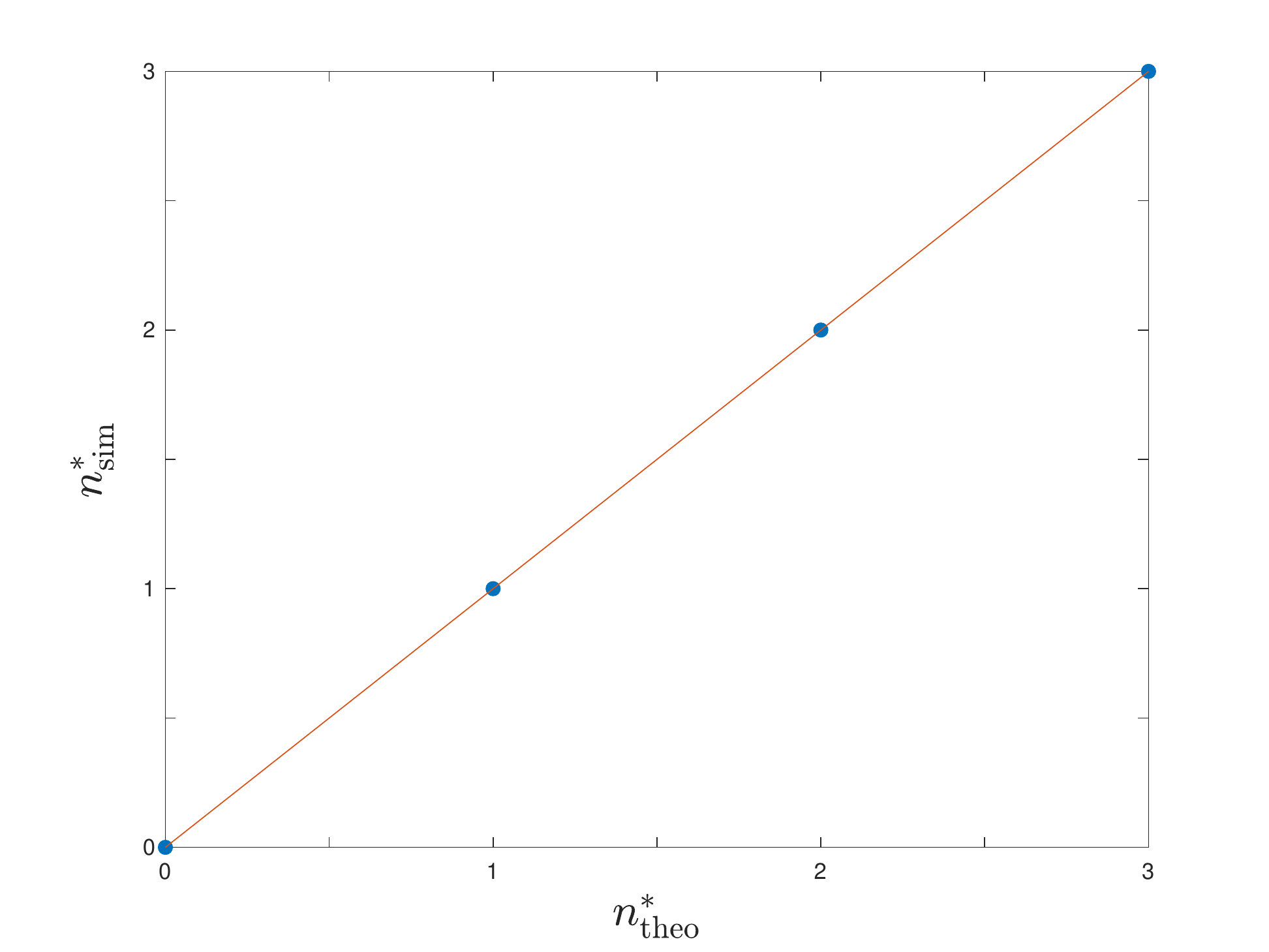}
\caption{\footnotesize\it Number of lock downs: comparison between analytical and numerical results obtained for an \ER random network with $N=10^5$ nodes, average degree $8$, $\beta_0=0$, $I_1=0.2$, $I_2=0.05$. The four points correspond to $\beta_1\in\{0.02,0.03,0.06,0.09\}$.}
\end{center}
\label{fig_SIR}
\end{figure}

\section{Inference of model parameters from epidemic data}
Here we consider the task of fitting against epidemic data a model including imperfect information on the state of the system.
This is an instance of a system identification problem \cite{keesman2011system}, which we solved along the following lines:
We consider the time series of observed new daily and active cases 
$(N_t^o, A_t^o)$ ($t=1\dots T$ is the temporal index in days, starting from the 1st of March, $T=T_l=155$ for Lombardy, and $T=T_b=100$ for the Basque country), and we assume it as coming from an instance of the model plus a noise term
\begin{eqnarray}
N_t^o = r I_u(t, \wp) + \delta_N \\
A_t^o = I_d(t, \wp) + \delta_A \\
\wp = \left\{ \gamma,\beta_1,\beta_0,r I_1, I_2, I_u(0), I_d(0) \right\},
\end{eqnarray}
where we highlighted the dependence of the model trajectory by the dynamical parameters and boundary values.

We assume shot-noise of the form  
\begin{eqnarray}
 \langle \delta_N \rangle = \langle \delta_A\rangle=0 \\
 \langle \delta_N^2 \rangle = N \quad \langle \delta_A^2 \rangle = A , 
\end{eqnarray} 
which for large numbers we assume to be distributed normally.
Upon assuming an uniform prior, we have the following formula for the log-likelihood of the parameters
\begin{equation}
\mathcal{L}(\wp) = \sum_t  \frac{(N_t^o - r I_u(t, \wp))^2}{2 N_t^o} + \frac{(A_t^o - I_d(t, \wp))^2}{2A_t^o}  + \textrm{const.}
\end{equation}
From the Bayes formula, the posterior probability distribution of parameters $P(\wp) \propto e^{-\mathcal{L}(\wp)}$ has been sampled by a Metropolis Montecarlo rule, where the evaluation of $\mathcal{L}(\wp)$ has been done by numerical integration of the model equations. More explicitly we have been following the following flowchart: 

\begin{itemize}
\item Start from some value of the parameters $\wp_0$: a warm start has been provided by fitting the curve of new daily cases alone in linear approximation ($S<<N$). 
\item Propose a change for the parameters $\wp_n \to \wp_{n+1}$: we used independent geometrical random walks of stepsize $10^{-3}$.
\item Numerically integrate the model equations with the new proposed parameters to evaluate their log-likelihood $\mathcal{L}(\wp_{n+1})$. We used the standard Verlet algorithm. 
\item Accept the proposed new parameters with probability \\
$\textrm{min}(1,\exp{ \mathcal{L}(\wp_{n+1}) - \mathcal{L}(\wp_{n})})$ (Metropolis rule), otherwise keep the old parameters.
\end{itemize}
This defines a series that asymptotically uniformly samples the posterior probability for the parameters, whose peak values have been used for the results showed in Fig. 6 and 7.  
Finally, we do point out that when we numerically integrate the model equations we rejected solutions that i) do not include at least one lock down event and ii) do not agree with prevalence estimate from serological data. The latter gives a lower bound for the total number $K$ of infected individuals at a certain time, $K\geq K_o$ that can be  reformulated as an inequality between the model parameters (in particular $r$, $\beta_1$ and $\beta_0$) as follows:
$K$ can decomposed in detected and undetected cases $K  = K_d + K_u$, where $K_d = \sum \Delta I_d^+$ is given by the data, while for the latter we have 
\begin{eqnarray}
K_u = \sum_t \Delta I_{u,t}^+ = \sum \beta(I_{d,t}) I_{u,t} = \\
=\frac{1}{r} \sum \beta(I_{d,t}) \Delta I_d^+ = \\
=\frac{\beta_1}{r} \sum_{incr} \Delta I_d^+ + \frac{\beta_0}{r} \sum_{decr} \Delta I_d^+ = \\
= \frac{K_d}{r} (\beta_1 (1-x) + \beta_0 x)
\end{eqnarray}
In the first and second lines we have used the model hypothesis, in the third we have decomposed the sum in terms of  the increasing and decreasing part of the wave, and finally in the last line $x$ is the fraction of the detected infections during the decreasing part of the epidemic wave.
We have finally the  inequality
\begin{equation}
r (K_o/K_d -1) \leq \beta_1 (1-x) + \beta_0 x
\end{equation}

Given the simplicity of the model employed -- in particular with respect to the hypothesis of a constant detection rate $r$ -- we obtain a fairly high value of the $\chi^2 \sim 20$, with a concomitant acceptably low average relative error of $\epsilon \sim 20\%$ across data points that can be considered apt for a qualitative description of the data.  
We report in the following table the maximum likelihood inferred parameters with their standard deviation.
   \begin{center}
\begin{tabular}{| l | l | l | l |}
\hline
Region &  $R_0 = \beta_1/\gamma$ & $1-p=\beta_0/\beta_1$ & $r$(day$^{-1})$  \\ \hline
Basque country  & $2.9\pm0.1$ & $ 1.3 \pm 0.4 \cdot 10^{-2}$ &  $3 \pm 1 \cdot 10^{-3}$ \\ \hline
Lombardy  & $5.2 \pm 0.2$ & $4 \pm 1 \cdot 10^{-2}$ & $8.4 \pm 0.6 \cdot 10^{-3}$ \\ \hline 
\end{tabular}
\end{center}

We do point out an anomalously high $R_0$ for Lombardy, due to a very low average recovery rate of approximately $1$ month (while the one from Basque country is around two weeks in agreement with WHO estimates). This is apparent upon looking at the much slower decay of the infection curves during the lock down, and it is probably due to a biased over-sampling of critically ill cases whose average recovery is typically longer. 

\bibliographystyle{unsrt}

\bibliography{ref_osc}

\end{document}